\documentclass[draft,12pt]{article}

\usepackage{amssymb, latexsym, amsmath, amsfonts}
\usepackage{amscd}
\usepackage{curves}
\usepackage{color}
\usepackage{amsthm}
\usepackage{amsbsy}

\def\bn{\begin{eqnarray}
 }
\def\en{\end{eqnarray}}

\newcommand{\half}{\tfrac{1}{2}}

\newcommand{\quarter}{\tfrac{1}{4}}

\def\s{\hskip.08em}
\def\b{\begin{eqnarray*}
 }
\def\e{\end{eqnarray*}}

\newcommand{\norm}[1]{\left\| #1 \right\|}
\newcommand{\abs}[1]{\left| #1 \right|}

\newenvironment{eq}{\begin{equation}}{\end{equation}}

\newcommand{\mat}[1]{\begin{pmatrix} #1 \end{pmatrix}}

\newcommand{\tr}{\operatorname{Tr}}

\renewcommand{\H}{{\mathcal{H}}}

\newcommand{\mbs}[1]{\makebox(0,0){\footnotesize #1}}
\newcommand{\qb}{\qbezier}

\begin{document}
\title{The distance between classical\\ and quantum systems}
\author{Deanna Abernethy and John R. Klauder\footnote{Also Physics 
Department.}\\
Department of Mathematics\\
University of Florida\\
Gainesville, FL  32611\\
Electronic mail: daber82@ufl.edu and klauder@phys.ufl.edu}

\date{}
\maketitle


In a recent paper, a ``distance" function, $\cal D$, was defined
which measures the distance between pure classical and quantum
systems. In this work, we present a new definition of a 
``distance",
$D$, which measures the distance between either pure or impure
classical and quantum states. We also compare the new distance
formula with the previous formula, when the latter is applicable. 
To
illustrate these distances, we have used $2 \times 2$ matrix
examples and 2-dimensional vectors for simplicity and clarity.
Several specific examples are calculated.

\section*{\textbf{1. INTRODUCTION}}

It is a property of nature that classical and quantum mechanics 
have
quite different characteristics.  Classical mechanics deals with
ordinary second order differential equations leading to particle
trajectories, while quantum mechanics involves the study of 
partial
differential equations for functions such as the wave function.
Despite this basic difference we wish to find a common 
formulation
of both classical and quantum mechanics.  We plan to use that 
common
formulation to define a distance between classical and quantum
systems.

Earlier work in this field, Ref. \cite{k1}, focused solely on 
pure
states, which were represented by vectors.  In units where 
Planck's
constant $\hbar = 1$, this distance formula was given by
\begin{eq}
        {\cal D}
       = \min_\alpha
               \int{
                       \|{
                               i\dot\psi(t) -
                               [{\dot\alpha} + {\H }]{\psi(t)}
                       }\|
               }dt.
       \label{calD}
\end{eq}
{}Here, we let $L \equiv i\dot\psi(t) - \bigl[\dot\alpha + \H
\bigr]\psi(t)$, where $\psi(t)$ is a Hilbert space vector and
$\norm{L}$ denotes the vector norm. If $\psi$ satisfies
Schr\"odinger's equation, then ${\cal D} = 0$.  However, if 
$\psi$
has a different temporal behavior, then generally ${\cal D} > 0$.

This formula for $\cal D$ corresponds to so-called {\it pure
states}, however there are other quantum states, namely {\it 
impure
states}, that cannot be covered by this original definition.

The purpose of the present work is to study some simple,
two-dimensional states and examine several examples of pure and
impure state situations.  We also verify that the results for the
pure states with the new formulation agree with the results 
obtained
by means of the previous formulation.  While these examples are 
only
for two-dimensional vectors and $2\times 2$ matrices, they
illustrate the general principles involved.

Our task is to find the distance between classical and quantum
systems.  At first glance, this seems comparable to finding the
distance between two dissimilar objects, such as a rock and a 
leaf.
Thus, we begin by trying to find a way to describe classical and
quantum systems in a similar manner in order that the two 
theories
can be compared.

The connection of classical and quantum mechanics used by Klauder 
in
forming his expression for the distance has been well discussed 
in
several papers; see Ref. \cite{k3}.  The essence of this 
connection
assumes the Schr\"odinger equation, but {\it restricts} the time
dependence of the wave function so it generally is \emph{not} a
solution to this equation.  When possible, this restricted time
dependence reflects the time dependence associated with classical
dynamics; the appropriate time dependence can be obtained by
restricting $\psi$ to evolve solely within coherent states for 
which
the parameters evolve as classical solutions.  However, for 
present
purposes, it suffices to choose a time dependence such that
$L\not\equiv0$. The deviation from $L$ being an exact solution is
then turned into an expression for the distance.

Klauder has proposed an alternative formula that should cover 
both
the pure and the impure states.  This new formula uses {\it 
density
matrices} $\rho$ and defines  a distance function $D$ by
\begin{eq}
       D = \int \norm{\,i\dot\rho(t) -
               \bigl[{\H },\rho(t) \bigr]\,} dt.
           \label{D}
\end{eq}
{}For ease of notation, we let $A \equiv i\dot\rho(t) - \bigl[{\H
},\rho(t)\bigr]$, where, in our examples, $\rho(t)$ is a 
Hermitian
$2\times 2$ matrix, $\dot\rho(t)$ is the time derivative of
$\rho(t)$, and $\H $ is the Hamiltonian $2\times 2$ matrix.  When
$A$ is a matrix, as is the case here, the symbol $\norm{A}$ 
denotes
the operator norm; for a discussion of operators and norms, see,
e.g., Ref. \cite{k2}.  Although the operator norm of $A$ is often
moderately difficult to evaluate, it is rather easy to calculate 
for
the $2\times 2$ matrices $A$, as we now demonstrate. It is 
because
of this ease of calculation that we focus our attention on 
matrices
of such a small dimension.

Let us note that $\rho$ and $\H $ are both Hermitian and that 
$\rho$
is normalized so that $\tr(\rho) = 1$. It then follows that $A$ 
is
anti-Hermitian and $\tr (A) = 0$. Thus, $A$ necessarily has the 
form
\begin{eq}
       A = U^\dag
               \mat{ d & \phantom{-}0 \\
           0 & -d } U,
\end{eq}
 where $U$ is a unitary $2\times 2$ matrix.

For our calculations we need to evaluate the usual operator norm 
of
$A$. The norm of $A$ is given by its largest eigenvalue and
therefore is $\abs{d}$. In order to determine this value, it is
useful for our examples to observe that
\begin{eq}  A^\dag A
       = U^\dag \mat{
               \abs{d}^2 & 0 \\
               0 & \abs{d}^2 }
               U.
\end{eq}
 Therefore
\begin{eq} \tr(A^\dag A) = 2\abs{d}^2. \end{eq}
Stated otherwise, it follows that
\begin{eq}
       \norm{A} = \sqrt{\half\, \tr(A^\dag A)}.
       \label{normA}
\end{eq}

It is important to note that this expression holds only when $A$ 
is
given by a $2\times 2$ matrix. For higher dimensional examples 
this
formula is generally inappropriate.

\section*{\textbf{2. GENERAL STRATEGY}}

As mentioned, the previous work on this subject dealt only with 
pure
states. We need to verify that the new distance formula 
accurately
measures the distance for pure states before we can move on to
impure states. We can accomplish this by comparing a pure state
example using our distance formula to the distance found in 
previous
work.  For convenience, we may find these examples as part of a 
more
general analysis.

We first start with the general form of a $2\times 2$ matrix, 
which
will lead to a general $\rho$.  This will allow us to apply the
general form to several, more specific pure and impure cases. Let
\begin{eq}
 \rho
       = \mat{a & b \\
           c & d }.
\end{eq}
However, $\rho$ is defined as a positive, Hermitian matrix with
$\tr(\rho) = 1$. This mandates that $a$ be real, $c = b^*$, the
complex conjugate of b, and $d = 1 - a$, as well as $0 \leq a 
\leq
1$. Thus, \begin{eq}
\rho = \mat{a & b\\ b^* & 1-a }.
\end{eq}

To differentiate between pure and impure states, we can appeal to
the value of $\tr(\rho^2)$.  A pure state has $\tr(\rho^2) = 1$,
while an impure state has $\tr(\rho^2) < 1$. Since

\begin{align}
\rho^2
%&= \mat{ a & b \\
%           b^* & 1-a }
 %          \mat{ a & b \\
  %         b^* & 1-a } \\
      &= \mat{a^2 + \abs{b}^2 &\ ab + b(1-a) \\
           ab^* + b^*(1-a) &\ \abs{b}^2 + (1-a)^2 },
\end{align}
we find that
\begin{align}
       \tr(\rho^2)         &= a^2 + 2\abs{b}^2 + (1-a)^2 \nonumber 
\\
                   &= 2a^2 + 2\abs{b}^2 + 1 - 2a.
\end{align}
To develop pure cases of $\rho$, we let $2a^2 + 2\abs{b}^2 + 1 - 
2a
= 1$, i.e.,
\begin{align}
      \abs{b} = \sqrt{a - a^2}.
\end{align}
Similarly, for impure cases, we set $\abs{b} < \sqrt{a - a^2}$.
Thus, for any choice of $a$, we can find a suitable value for 
$b$.
We next apply this formula to create several examples.

\subsubsection*{General example 1:}

We will be complete in our calculations for this example, but
certain steps will be omitted in further calculations for
simplicity.

Let us begin by choosing $a = \cos^2(t)$, then
\begin{align}
       \abs{b}
       &= \bigl[\cos^2(t) - \cos^4(t) \bigr]^{1/2} \nonumber \\
          &= \half \abs{\sin(2t)}.
\end{align}
Therefore, we may set $b = \half \sin(2t)$ for pure cases, and
$\abs{\,b} < \half \abs{\,\sin(2t)}$ for impure cases. For 
example,
if we choose $b =\quarter \sin(2t)$, we have satisfied this
condition for an impure case, for general $t$ values.

Notice that the two cases differ by just a coefficient $\beta$. 
If
we keep the same value for $a$, but allow $b = \beta \sin(2t)$ 
with
$\abs{\beta} \leq \half$, then we can evaluate both pure ($\beta=
\half$) and selected impure cases ($\beta = \quarter$ and $\beta 
=
0$) from this general matrix. Thus we adopt $\rho$ and $\H $ for 
our
first set of examples as
\begin{eq} \rho = \mat{\cos^2(t)  &\beta\sin(2t) \\
           \beta\sin(2t)  &\sin^2(t) }
\end{eq}
and
\begin{eq} \H  = \lambda
            \mat{
           1 & \phantom{-}0 \\
           0 & -1}.
\end{eq}
To evaluate the distance $D$ (defined in Eq.~\ref{D}), we begin 
by
finding $i\dot\rho(t)$ as given by
\begin{eq} i\dot\rho(t)
       = i \mat{-2\cos(t)\s\sin(t) & 2\beta\cos(2t) \\
               2\beta\cos(2t) & 2\sin(t)\s\cos(t) }
\end{eq}
and $\bigl[{\H }, \rho(t)\bigr]$ as
\begin{align}
\bigl[\H , \rho(t)\bigr]
%        &=        \lambda \mat{1 & 0 \\
%               0 & -1 }
 %          \mat{\cos^2(t) & \beta\,\sin(2t) \\
  %             \beta\,\sin(2t) & \sin^2(t)} \\
   %  &\hskip60pt
    %    - \lambda \mat{\cos^2(t) & \beta\s\sin(2t) \\
     %          \beta\s\sin(2t) & \sin^2(t) }
      %         \mat{1 & 0 \\
       %        0 & -1 } \\
   &= \lambda \mat{0 & 2\beta\s\sin(2t) \\
               -2\beta\s\sin(2t) & 0 }.
\end{align}
This allows us to calculate the value of $A$ as

%**********************************

\begin{align}
A  &= \mat{ -2i\,\cos(t)\,\sin(t)
                          &\ 2i\,\beta\,\cos(2t) -
                       2\lambda\,\beta\,\sin(2t) \\
       2i\,\beta\,\cos(2t) + 2\lambda\,\beta\s\sin(2t)
               &\ 2i\sin(t)\s\cos(t) }.
\end{align}
As follows from Eq. \eqref{normA}, $\norm{A}$ is given by
\begin{align}
\norm{A}
       &= \bigl[ \half \tr(A^\dag A)
               \bigr]^{1/2} \nonumber \\
       &= \bigl[
           4\cos^2(t)\s\sin^2(t)
           \ +\ 4\beta^2\s\cos^2(2t)
           \ +\ 4\lambda^2\s\beta^2\s\sin^2(2t)
               \bigr]^{1/2}.
\label{first}
\end{align}
This gives us a guideline for computing the distance for this 
$\rho$
and this $\H $.

Now we can confirm our use of the distance formula $D$ by 
comparing
our distance for a pure state to the distance using the method
previously described in Ref. \cite{k1}.

\subsubsection*{Example 1a (pure case, $\beta=\half$):}

Using this general sample case, we begin with a pure case of 
$\rho$
by choosing $a = \cos^2(t)$ and $b = \half\sin(2t) = b^*$. Thus
\begin{eq} \rho = \mat{ \cos^2 (t) & \frac{1}{2}\,\sin (2t)  \\
          \frac{1}{2}\,\sin (2t)  &  \sin^2 (t) }
\end{eq}
and
\begin{eq} \H  = \lambda \mat{1 & \phantom{-}0 \\ 0 & -1}.
\end{eq}
If
we apply $\beta = \half$ to Eq. \eqref{first}, we get
\begin{align}
\norm{A}
       &= \bigl[ 4\cos^2(t)\s\sin^2(t) + \cos^2(2t)
               +\lambda^2\sin^2(2t) \bigr]^{1/2} \nonumber \\
       &= \bigl[ 1 + 4\lambda^2\cos^2(t)\s\sin^2(t)
               \bigr]^{1/2}.
                       \label{1a}
\end{align}
Figure 1 illustrates the equation for $\norm{A}$ in this example 
for
$\lambda=1$.

%--------------------------
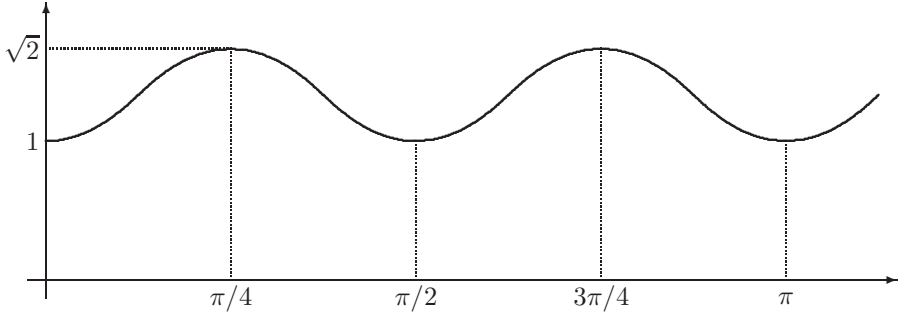
\begin{figure}
\setlength{\unitlength}{.7pt}
\begin{center}
\begin{picture}(475,160)(-15,-110)

\put(0,-110){\vector(0,1){160}} \put(-10,-100){\vector(1,0){470}}

\multiput(0,-25)(200,0){2}{%
       \thicklines
       \qb(0,0)(25,0)(50,25)
       \qb(50,25)(75,50)(100,50)
       \qb(100,50)(125,50)(150,25)
       \qb(150,25)(175,0)(200,0)
       \thinlines
       \qb[38](200,0)(200,-37.5)(200,-75)
       \qb[63](100,50)(100,-12.5)(100,-75)
} \put(400,-25){\thicklines\qb(0,0)(25,0)(50,25)}

\put(100,-110){\mbs{$\pi/4$}} \put(200,-110){\mbs{$\pi/2$}}
\put(300,-110){\mbs{$3\pi/4$}} \put(400,-110){\mbs{$\pi$}}

\put(-7,-25){\mbs{$1$}} \put(-13,25){\mbs{$\sqrt{2}$}}

\qb[50](0,25)(50,25)(100,25)

\end{picture}
\end{center}
\setlength{\unitlength}{1pt} \label{pic1} \caption{Graph of
$\norm{A}$ {\it vs.} $t$ for Example 1a}
\end{figure}
%------------------------------

\subsection*{Comparison of the Two Distance Functions}

Before proceeding further, let us verify that the new distance
function $D$ agrees with the distance $\cal D$ using Eq.
\eqref{calD}. We begin by identifying $\psi(t)$ as follows
\begin{eq}
        \psi(t) = \mat{ \cos(t) \\  \sin(t) }.
\end{eq}
%We also calculate $i\dot\psi(t)$ as
%\[ i\dot\psi(t)
%= i\mat{ -\sin(t) \\ \phantom {-}\cos(t) }
%                = \mat{ -i\sin(t) \\ \phantom{-}i\cos(t) } \]\;.
%Next we have $\bigl[ \dot\alpha + {\H } \bigr]{\psi(t)}$ given 
by
%\begin{align*}
%\bigl[ \dot\alpha + {\H } \bigr]{\psi(t)}
%        &= \bigg[\dot\alpha \ +\  \lambda
%               \mat{ 1 & \phantom{-}0 \\
 %      0 & -1} \bigg]
  %     \mat{ \cos(t) \\ \sin(t) } \\
%       &= \mat{ (\dot\alpha + \lambda)\cos(t) \\
 %      (\dot\alpha - \lambda)\sin(t) }.
%\end{align*}
It follows that the integrand in Eq. \eqref{calD} involves
\begin{align}
L = i\dot\psi(t) - \bigl[ \dot\alpha + {\H } \bigr]{\psi(t)}
%       &= \mat{ -i\sin(t) \\
 %          \phantom{-}i\cos(t) }
  %         \ -\
   %        \mat{ (\dot\alpha + \lambda)\cos(t) \\
    %       (\dot\alpha - \lambda)\sin(t) } \\
       &= \mat{ -i\sin(t) - (\dot\alpha + \lambda)\cos(t) \\
   \phantom{-}i\cos(t) - (\dot\alpha - \lambda)\sin(t) }\;,
\end{align}
and the vector norm of this expression is
\begin{align}
       \mbox{}\hskip50pt\mbox{}&
       \hskip-50pt
       \|{i\dot\psi(t) -
       [ \dot\alpha + {\H } ]{\psi(t)} }\|
               \notag \\
  %     &= \bigl[ \ \abs{-i\sin(t)
   %                    - (\dot\alpha + \lambda)\cos(t)}^2
    %           \ +\  \abs{i\cos(t)
     %                  - (\dot\alpha -\lambda)\sin(t)}^2 \
      %         \bigr]^{1/2}
       %                \nonumber \\
              &= \bigl[ \sin^2(t)
               \,+\, (\dot\alpha + \lambda)^2\cos^2(t)
               \,+\, \cos^2(t)
               \,+\,(\dot\alpha - \lambda)^2\sin^2(t)
               \bigr]^{1/2}
                       \nonumber\\
       &= \bigl[ 1 \,+\, \dot\alpha^2 \,+\, \lambda^2
               \,+\, 2\dot\alpha\lambda\cos(2t) \bigr]^{1/2}.
                       \label{psi1a}
\end{align}
We can minimize $\norm{L}$ over ${\dot\alpha}$ by choosing
\begin{eq}
    \dot\alpha
       \,=\,  -\lambda\cos(2t).
\end{eq}
Substituting this value for $\dot\alpha$ into Eq. \eqref{psi1a}, 
we
obtain
\begin{align}
%     &=\bigl[1 + \dot\alpha^2
%       +\lambda^2
 %              + 2\dot\alpha\lambda\cos(2t) \bigr]^{1/2}\\
     \norm{L}  &= \bigl\{
               1
               + \bigl[-\lambda\cos(2t)\bigr]^2 + \lambda^2
               + 2\bigl[-\lambda\cos(2t)\bigr]\lambda\cos(2t)
               \bigr\}^{1/2} \nonumber \\
       &= \bigl[
               1 + 4\lambda^2\cos^2(t)\s\sin^2(t)
               \bigr]^{1/2}\;,
\end{align}
a result that agrees with Eq. \eqref {1a}. Thus our definition of
the new distance function $D$ is appropriate, and we will 
continue
to apply this formula to further examples.

\section*{\textbf{3. ADDITIONAL SPECIFIC CASES}}

\subsubsection*{Example 1b (impure case, $\beta=\quarter$):}

Resuming our calculations, we will evaluate an impure case with $a 
=
\cos^2(t)$ and $b = \quarter\sin(2t)$. This gives
\begin{eq} \rho =
       \mat{ \cos^2(t) & \quarter\sin(2t) \\
           \quarter\sin(2t) & \sin^2(t) }
\end{eq}
and
\begin{eq} \H  = \lambda
               \mat{ 1 & 0 \\
       0 & -1 }.
\end{eq}
Again using equation \eqref{first}, we obtain a value for 
$\norm{A}$
as
\begin{align}
\norm{A}
       &= \bigl[4\cos^2(t)\s\sin^2(t) + \quarter\cos^2(2t)
               +\quarter\lambda^2\sin^2(2t)\bigr]^{1/2} \nonumber 
\\
       &= \bigl[\quarter
               + (3 + \lambda^2)\cos^2(t)\s\sin^2(t)\bigr]^{1/2}.
\end{align}

\subsubsection*{Example 1c (impure case, $\beta = 0$):}

We will now choose $b = 0$, which also gives an impure case. We 
have
\begin{eq} \rho =\mat{\cos^2(t) & 0 \\0 & \sin^2(t)}
\end{eq}
and $\H $ is still given by
\begin{eq} \H =\lambda \mat{1 & \phantom{-}0 \\
0 & -1 }.
\end{eq}
We evaluate $\norm{A}$ as
\begin{align}
\norm{A}
       &= [4\cos^2(t)\s\sin^2(t)]^{1/2} \nonumber \\
       &= \abs{\sin(2t)}.
\end{align}
In this case, when we evaluate the distance, say for $T = 4\pi$, 
we
are led to
\begin{align}
D
       &= \int_0^{4\pi}
               \abs{\sin(2t)} dt \nonumber \\
       &= 8\int_0^{\pi/2} \sin(2t) dt \nonumber \\
       &= 8.
\end{align}

\subsubsection*{General example 2:}

With $a = \cos^2(t)$, we again have $b = \beta \sin(2t) = b^*$ 
for
$\abs{\beta} \leq {\half}$ to give us
\begin{eq} \rho(t) = \mat{
   \cos^2(t) & \beta\sin(2t) \\
   \beta\sin(2t) & \sin^2(t)}\;,
\end{eq}
but now we choose a new expression for $\H$ given by
\begin{eq} \H
= \lambda \mat{
           0 & 1 \\
           1 & 0 }.
\end{eq}
It follows that
\begin{align}
A &= i{\dot\rho}(t)-[\rho(t),\H] \nonumber \\
%       &= i \mat{ -\sin(2t) & 2\beta\cos(2t) \\
 %      2\beta\cos(2t) & \sin(2t) }
%   \ - \ \lambda \mat{
 %      0 & -\cos(2t) \\
  %     \cos(2t) & 0 } \\
       &= \mat{ -i\sin(2t) & 2i\beta\cos(2t) + \lambda\cos(2t) \\
       2i\beta\cos(2t) - \lambda\cos(2t) & i\sin(2t) }.
\end{align}

Hence
\begin{eq}
A^\dag A = \mat{
       j_1 & j_2 \\
       j_3 & j_4 },
\end{eq}
where
\begin{align}
j_1
       &= \sin^2(t) \ +\ \bigl[2i\beta\cos(2t) -
               \lambda\cos(2t)\bigr]\bigl[-2i\beta\cos(2t) -
               \lambda\cos(2t)\bigr] \\
j_2
       &= \bigl[i\sin(2t)\bigr]\bigl[2i\beta\cos(2t) +
               \lambda\cos(2t) \bigr] \nonumber \\
               &\hskip50pt +\ \bigl[i\sin(2t)\bigr]\bigl[
               -2i\beta\cos(2t) - \lambda\cos(2t) \bigr] \\
j_3
       &= \bigl[-i\sin(2t)\bigr]\bigl[-2i\beta\cos(2t) +
               \lambda\cos(2t)\bigr] \nonumber \\
               &\hskip50pt -\ \bigl[i\sin(2t)\bigr]
               \bigr[2i\beta\cos(2t) - \lambda\cos(2t)\bigr] \\
j_4
       &= \bigl[-2i\beta\cos(2t) +
               \lambda\cos(2t)\bigr]\bigl[2i\beta\cos(2t) -
               \lambda\cos(2t)\bigr]\ +\ \sin^2(2t)\;,
\end{align}
which leads to
\begin{align} \norm{A}
       &= \sqrt{\half \tr(A^\dag A)} \nonumber \\
       &= \tfrac{1}{\sqrt{2}}\bigl[2\sin^2(2t) + 
4\beta^2\cos^2(2t)
               + \lambda^2\cos^2(2t) \nonumber \\
               &\hskip50pt + 4\beta^2\cos^2(2t)
               + \lambda^2\cos^2(2t)\bigr]^{1/2} \nonumber \\
       &= \bigl[\sin^2(2t) + \cos^2(2t)(4\beta^2
               + \lambda^2)\bigr]^{1/2}.
\label{second}
\end{align}
Just as in Example 1, we can choose values of $\beta$ to create 
both
pure and impure cases.

\subsubsection*{Example 2a (pure case, $\beta = \half$):}

Our pure choice of $\beta = \half$ gives $b = \half \sin(2t) = 
b^*$
and $\rho(t)$ as follows
\begin{eq}
\rho(t)
       = \mat{ \cos^2(t) & \half\sin(2t) \\
           \half\sin(2t) & \sin^2(t) }
\end{eq}
and
\begin{eq} \H
       = \lambda \mat{ 0 & 1 \\
   1 & 0 }.
\end{eq}
By following Eq. \eqref{second}, we get the value for $\norm{A}$
when $\beta = \half$ given by
\begin{align}
\norm{A}
%        &= \bigl[\sin^2(2t) + \cos^2(2t)\bigl(4(\quarter)^2
%               + \lambda^2\bigr)\bigr]^{1/2} \\
       &= \bigl[\sin^2(2t) + \cos^2(2t)(1
               + \lambda^2)\bigr]^{1/2}.
\end{align}

\subsubsection*{Example 2b (impure case, $\beta = \quarter$):}

In this case
\begin{eq}
\rho(t) = \mat{ \cos^2(t) & \quarter\sin(2t) \\
                   \quarter\sin(2t) & \sin^2(t) }
\end{eq}
and
\begin{eq} \H  = \lambda \mat{ 0 & 1 \\
       1 & 0 }.
\end{eq}
With the aid of Eq. \eqref{second}, we get
\begin{align}
\norm{A}
%        &= \bigl\{\sin^2(2t) + \cos^2(2t)[(4)(\quarter)^2
%               + \lambda^2]\bigr\}^{1/2} \\
       &= \bigl[\sin^2(2t) + \cos^2(2t)(\quarter
               + \lambda^2)\bigr]^{1/2}.
\end{align}

\subsubsection*{Example 2c (impure case, $\beta = 0$):}

Our final impure case is $\beta = 0$.  This gives us
\begin{eq}
\rho(t) = \mat{ \cos^2(t) & 0 \\
       0 & \sin^2(t) }
\end{eq}
and
\begin{eq}
\H  = \lambda \mat{ 0 & 1 \\
   1 & 0 }.
\end{eq}
Following Eq. \eqref{second} once more, we get
\begin{eq} \norm{A}
       = \bigl[\sin^2(2t) + \lambda^2\cos^2(2t)\bigr]^{1/2}.
\end{eq}

We have investigated several impure and pure cases with two
different choices of $\H $.  Next, we will explore new general
examples of
$\rho(t)$ with the same two choices of $\H $. \\

\subsubsection*{General example 3:}

We will now obtain a new general $\rho(t)$ by choosing $a = 1/(1 
+
t^2)$.  Then the limiting $b$ is found in the following manner
\begin{align} \abs{b}
       &= \sqrt{a - a^2} \nonumber \\
   &= \bigg\{\frac{1}{1 + t^2} -\bigg[\frac{1}{1
           + t^2}\bigg]^2\bigg\}^{1/2} \nonumber \\
   &= \frac{\abs{t}}{1 + t^2}.
\end{align}
Thus, for a pure case, we can pick $b = t/(1 + t^2)$, and $|b| <
\abs{t}/(1 + t^2)$ for impure cases. As was previously the case 
we
let the pure and impure choices of $b$ differ only by a real
constant $\beta$.  We can evaluate a general case of $b = \beta 
t/(1
+ t^2)$ with $\abs{\beta} \leq 1$ to enable us to consider 
several
specific cases. Thus we are led to consider
\begin{eq}
\rho(t)
       = \frac{1}{1 + t^2} \mat{ 1 & \beta t \\
   \beta t & t^2 }\;,
\end{eq}
while initially we also use our original $\H $, namely,
\begin{eq}
\H
       = \lambda \mat{ 1 & \phantom{-}0 \\
   0 & -1 }.
\end{eq}

We find $A=i{\dot\rho}(t)-[\H,\rho(t)]$ is given by
\begin{align}
A
%        &= \frac{i}{[1 + t^2]^2} \mat{ -2t & \beta(1 - t^2) \\
%       \beta(1 - t^2) & 2t }
 %  \ -\ \frac{\lambda}{1 + t^2} \mat{ 0 & 2\beta t \\
  %     -2\beta t & 0 } \\
       &= \frac{1}{[1 + t^2]^2} \mat{ -2it & i\beta (1 - t^2)
                       - 2\lambda\beta t(1 + t^2)  \\
       i\beta (1 - t^2) + 2\lambda\beta t(1 + t^2) & 2it }\;,
\end{align}
while our calculation of $\norm{A} = \sqrt{\half \tr(A^\dag A)}$
gives
\begin{align} \norm{A}
%        &= \bigg\{ \frac{1}{[2(1 + t^2)]^4} \bigl[8t^2 + \beta^2 
(1           %      - t^2) \nonumber \\
%       &\hskip50pt +\ 4\lambda^2\beta^2 t^2 (1 + t^2)^2 + \beta^2 
(1          %       - t^2)^2 + 4\lambda^2\beta^2 t^2 (1
 %              + t^2)^2\bigr]\bigg\}^{1/2} \nonumber \\
       &= \frac{\{4t^2 + \beta^2 (1 - t^2)^2
               + 4\lambda^2\beta^2 t^2 (1 + t^2)^2\}^{1/2}}{(1 +
               t^2)^2} \label{third}.
\end{align}

\subsubsection*{Example 3a (pure case, $\beta = 1$):}

Let us choose $\beta = 1$, which means that
\begin{eq}
\rho(t)
       = \frac{1}{1 + t^2} \mat{ 1 & t \\
   t & t^2 }
\end{eq}
with the same $\H $, i.e.,
\begin{eq} \H
       = \lambda \mat{ 1 & 0 \\
       0 & -1 }.
\end{eq}
Applying Eq. \eqref{third}, we find that
\begin{align}
\norm{A}
%        &= \frac{[4t^2 + (1 - t^2)^2 + 4\lambda^2 t^2(1
%               + t^2)^2]^{1/2}}{(1 + t^2)^2} \notag \\
       &= \frac{[1 + 4\lambda^2 t^2]^{1/2}}{1 + t^2}.
\label{3a}
\end{align}
Figure 2 is an illustration of the equation for $\norm{A}$ in 
this
example for $\lambda=1$.

%-------------------------------------------
\begin{figure}
\setlength{\unitlength}{1.5pt}
\begin{center}
\begin{picture}(230,100)(-10,-10)

\put(-5,0){\vector(1,0){220}} \put(0,-5){\vector(0,1){95}}

\thicklines \qb(0,50)(15,50)(30,60) \qb(30,60)(45,70)(60,70)
\qb(60,70)(75,70)(95,60) \qb(95,60)(120,47)(200,35)

\thinlines \qb[60](0,70)(30,70)(60,70) 
\qb[70](60,70)(60,35)(60,0)

\put(-13,70){\mbs{$2\sqrt{3}\big/3$}} \put(-5,50){\mbs{$1$}}
\put(60,-7){\mbs{$\sqrt{2}\big/2$}}

\end{picture}
\end{center}
\setlength{\unitlength}{1pt} \label{pic2} \caption{Graph of
$\norm{A}$ {\it vs.} $t$ for Example 3a}
\end{figure}
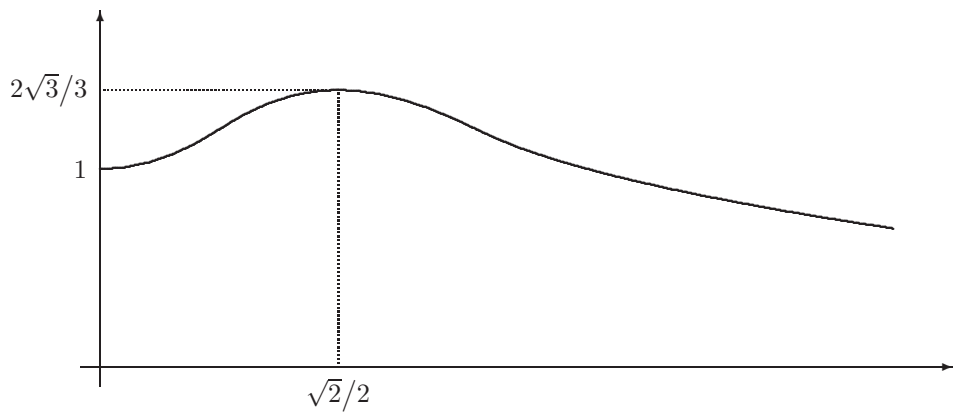
%----------------------------------------------

\section*{\textbf{4. ANOTHER COMPARISON OF THE\\ TWO DISTANCE 
FUNCTIONS}}

Let us again verify that our definition of $D$ is suitable by
evaluating $\cal D$ for the information in Example 3a and 
comparing
it to Eq. \eqref{3a}.

We begin by identifying
\begin{eq} \psi(t)
   =   \frac{1}{\sqrt{1 + t^2}} \mat{ 1 \\ t }\;,
\end{eq}
which leads directly to
\begin{align}
L= i\dot\psi(t) - [\dot\alpha + \H]\psi(t)
%    &=  \frac{i}{(1 + t^2)^{3/2}} \mat{ -t \\ 1}
%       \ -\ \frac{1}{(1 + t^2)^{1/2}} \mat{
 %      \dot\alpha + \lambda \\
  %     t(\dot\alpha - \lambda) } \\
   &=  \frac{1}{(1 + t^2)^{3/2}} \mat{
       -it - (\dot\alpha + \lambda)\ (1 + t^2) \\
       i - t(\dot\alpha - \lambda)\ (1 + t^2) }.
\end{align}
Finally, $\norm{L}=\|{i\dot\psi(t) - [\dot\alpha + \H]\psi(t)}\|$ 
is
\begin{align}
\norm{L}
%    &=  \frac{\bigl[t^2 + (\dot\alpha + \lambda)^2\ (1 + t^2)^2 + 
1 +
%       t^2(\dot\alpha - \lambda)^2\ (1 + t^2)^2 \bigr]^{1/2}}{(1 
+
 %      t^2)^{3/2}} \notag \\
   &=  \frac{\bigl[\dot\alpha^2(t^2 + 1)^2 -
       2\lambda\dot\alpha(t^2 - 1)(t^2 + 1) + \lambda^2(t^2 + 1)^2 
+
       1\bigr]^{1/2}}{1 + t^2}.
\label{psi3a}
\end{align}
We next minimize $\norm{L}$ by choosing
\begin{eq} \dot\alpha
   =   \lambda \bigg(\frac{t^2 - 1}{t^2 + 1}\bigg).
\end{eq}
When we substitute this choice back into Eq. \eqref{psi3a} we 
find
that
\begin{eq}
\frac{\bigl[\dot\alpha^2(t^2 + 1)^2 -
       2\lambda\dot\alpha(t^2 - 1)(t^2 + 1) + \lambda^2(t^2 + 1)^2 
+
       1\bigr]^{1/2}}{1 + t^2} \\
   =   \frac{[1+4\lambda^2 t^2]^{1/2}}{1+t^2 },
\end{eq}
a result that coincides with Eq. \eqref{3a}.

\section*{\textbf{5. CONTINUATION OF SPECIFIC CASES}}

\subsubsection*{Example 3b (impure case, $\beta = \half$):}

By referring to Eq. \eqref{third}, we can also evaluate 
$\norm{A}$
when $\beta = \half$ as
\begin{align}
\norm{A}
%    &=   \frac{\bigl[4t^2 + \quarter (1 - t^2)^2 + \lambda^2 t^2 
(1 +
%       t^2)^2\bigr]^{1/2}}{(1 + t^2)^2} \\
   &=  \frac{[(t^2 + 1)^2 (\quarter + \lambda^2 t^2) +
       3t^2]^{1/2}}{(1 + t^2)^2}.
\end{align}

\subsubsection*{Example 3c (impure case, $\beta = 0$):}

Similarly, we can substitute an impure case of $\beta = 0$ into 
Eq.
\eqref{third} to find
\begin{align}
\norm{A}
%   &=   \frac{\sqrt{4t^2}}{(1 + t^2)^2} \\
   &=  \frac{2|t|}{(1 + t^2)^2}.
\end{align}

\subsubsection*{General example 4:}

We will use the same value $a = 1/(1 + t^2)$ with $b = (\beta 
t)/(1
+ t^2)$ for $\abs{\beta} \leq 1$. As before, this gives
\begin{eq}
\rho(t)
       = \frac{1}{1 + t^2} \mat{ 1 & \beta t \\
   \beta t & t^2 }\;,
\end{eq}
but now we use a new expression of $\H$ given by
\begin{eq}
\H
       = \lambda \mat{ 0 & 1 \\
   1 & 0 }.
\end{eq}

It follows that $A=i{\dot{\rho}}(t)-[\H,\rho(t)]$ is given by
\begin{align} A
%       &= \frac{i}{(1 + t^2)^2} \mat{ -2t & \beta (1 - t^2) \\
 %          \beta (1 - t^2) & 2t }\ -\ \frac{\lambda}{1
  %             + t^2} \mat{ 0 & t^2 - 1 \\
   %    1 - t^2 & 0 } \\
       &= \frac{1}{(1 + t^2)^2} \mat{ -2it & i\beta (1 - t^2)
               - \lambda (t^4 - 1)  \\
       i\beta (1 - t^2) + \lambda (t^4 - 1) & 2it }\;,
\end{align}
and so $\norm{A} = \sqrt{{\half} \tr(A^\dag A)}$ is
\begin{align}
\norm{A}
%       &= \bigg\{\frac{1}{2(1 + t^2)^4} \bigl[8t^2 + \beta^2 (1
 %              - t^2)^2 + \lambda^2 (t^4 - 1)^2 + \beta^2 (1 - 
t^2)^2           %      \nonumber \\
  %     &\hskip50pt +\ \lambda^2 (t^4
    %           - 1)^2 \bigr]\bigg\}^{1/2} \nonumber               
  \\
       &= \frac{[4t^2 + \beta^2 (1 - t^2)^2 + \lambda^2 (t^4
               - 1)^2]^{1/2}}{(1 + t^2)^2}.
\label{fourth}
\end{align}

\subsubsection*{Example 4a (pure case, $\beta = 1$):}

We first choose $\beta = 1$ to create a pure state case, namely
\begin{eq}
\rho(t)
       = \frac{1}{1 + t^2} \mat{ 1 & t \\
       t & t^2 }\;.
\end{eq}
With $\beta=1$, it follows from Eq. \eqref{fourth} that
\begin{align}
\norm{A}
%        &= \frac{[4t^2 + (1 - t^2)^2 + \lambda^2 (t^4
%               - 1)^2]^{1/2}}{(1 + t^2)^2} \\
       &= \frac{[1 + \lambda^2 (t^2 - 1)^2]^{1/2}}{1 + t^2}.
\end{align}

\subsubsection*{Example 4b (impure case, $\beta = \half$):}

We can investigate an impure case if we choose $\beta = \half$ 
and
substitute it into Eq. \eqref{fourth} as given by
\begin{align}
\norm{A}
%    &=   \frac{[4t^2 + \quarter (1 - t^2)^2 + \lambda^2 (t^4 -
%       1)^2]^{1/2}}{(1 + t^2)^2} \\
   &=  \frac{[\quarter (t^2 + 1)^2 + 3t^2 + \lambda^2 (t^4 -
       1)^2]^{1/2}}{(1 + t^2)^2}.
\end{align}

\subsubsection*{Example 4c (impure case, $\beta = 0$):}

Finally, when we choose an impure case with $\beta = 0$, we find
that
\begin{eq}
\norm{A}
   =   \frac{[4t^2 + \lambda^2 (t^4 - 1)^2]^{1/2}}{(1 + t^2)^2}.
\end{eq}

\section*{\textbf{6. SUMMARY}}

In this article we have introduced  a ``distance function", 
defined
by Eq. \eqref{D}, which measures the distance between classical 
and
quantum states, either pure or impure.  We have verified that the
previous method, defined in Eq. \eqref{calD}, and the new 
technique
for calculating distance are equivalent in the case of pure 
states.
We then found the distance for several combinations of density
matrices, $\rho(t)$, and Hamiltonians, $\H $.

To clearly illustrate the basic ideas, we have limited our 
interest
to $2\times 2$ density matrices. Additional studies for larger
density matrices, either finite or infinite dimensional, would 
also
be of interest. This is especially true for examples based on
coherent states which are appropriate to discuss traditional
classical situations; see Ref. [1] for such an analysis for the 
pure
state case.

\section*{\textbf{7. ACKNOWLEDGMENTS}}

We thank  Dr. Alexandru Scorpan for his assistance in preparing 
the
graphs. The present article is based on the Senior Honors Thesis
presented to the Department of Mathematics, University of 
Florida,
April, 2004, by one of us (DA).

\end{document}